\documentclass{JHEP3} % 10pt is ignored!

\usepackage{epsfig,multicol}
\usepackage{amsmath, amssymb}

\usepackage{concmath, palatino}

\newcommand\fverb{\setbox\pippobox=\hbox\bgroup\verb}
\newcommand\fverbdo{\egroup\medskip\noindent%
			\fbox{\unhbox\pippobox}\ }
\newcommand\fverbit{\egroup\item[\fbox{\unhbox\pippobox}]}
\newbox\pippobox
\newcommand{\be}{\begin{equation}} 
\newcommand{\ee}{\end{equation}}

\newcommand{\ba}{\begin{eqnarray}}
\newcommand{\ea}{\end{eqnarray}}

\newcommand{\zz}{\mathbb{Z}}

\newcommand{\ads}{AdS_5\times S^5}

\newcommand{\refeq}[1]{Eq.~(\ref{#1})}

\title{Anomalous dimensions at twist-3 in the $\mathfrak{sl}(2)$ sector of ${\cal N}=4$ SYM}

\author{Matteo Beccaria\\
  Dipartimento di Fisica, Universita' di Lecce,
  Via Arnesano, 73100 Lecce\\
  INFN, Sezione di Lecce\\
  E-mail: \email{matteo.beccaria@le.infn.it}}

\preprint{}

\abstract{
We consider twist-3 operators in the $\mathfrak{sl}(2)$ sector of ${\cal N}=4$ SYM built out 
of three scalar fields with derivatives. We extract from the Bethe Ansatz equations of this sector
the exact lowest anomalous dimension $\gamma(s)$ of scaling fields for several values of the operator spin $s$. 
We propose compact closed expressions for the spin dependence of $\gamma(s)$ up to the four loop level and show that 
they obey a simple new {\em twist-3 transcendentality principle}. As a check, we reproduce the four loop universal 
cusp anomalous dimension governing the logarithmic large spin limit of $\gamma(s)$.

\vskip 2cm
\centerline{\em to the memory of Giuseppe Curci}
}

\begin{document}

\section{Introduction}
\label{sec:Intro}

Integrability in QCD is an old intriguing issue (see~\cite{Belitsky:2004cz} for a recent review). 
Certainly, QCD is not an integrable quantum field theory.
However, in the planar multicolor 't Hooft limit, new special features emerge due to the simplified dynamics.
For instance, it is known that the one-loop renormalization group flow of 
certain sets of composite operators is associated with integrable XXX lattice Hamiltonians 
with $\mathfrak{sl}(2)$ spin symmetry~\cite{Lipatov:1993yb}. 
The evolution time in the integrable model is the (logarithm of the) renormalization scale at which 
we define the composite operators. The integrable Hamiltonian is identified with the quantum dilatation operator. 
Its eigenvalues are the anomalous dimensions of the scaling fields. Several explicit examples can be found 
in~\cite{Braun:1998id,Braun:1999te,Belitsky:1999qh,Belitsky:1999ru,Belitsky:1999bf,Derkachov:1999ze,Beisert:2004fv}.

\medskip
Inspired by the parallel developments of AdS/CFT duality~\cite{Maldacena:1997re},
the integrability properties of large $N$ gauge theories 
have been deeply investigated in supersymmetric Yang-Mills theories. In this paper, we shall work in the context 
of the maximally supersymmetric ${\cal N}=4$ super Yang-Mills theory which is UV finite and superconformally invariant at the quantum level.
In the analysis, we shall always understand the planar limit.
Due to the possible interdisciplinary interest of our investigation, we briefly overview the main 
logical developments of integrability in ${\cal N}=4$ SYM.

\medskip
The first positive results  are described in the seminal paper~\cite{Minahan:2002ve}. The one loop
dilatation operator is computed in a closed $\mathfrak{so}(6)$ sector of scalar operators. It can be identified with the
Hamiltonian of an integrable lattice model. Soon, the integrable structure could be extended to the full
set of $\mathfrak{psu}(2,2|4)$ operators. Also, it appeared clear that integrability could work beyond one loop, 
with the coupling constant being a deformation parameter for the superconformal algebra 
representation~\cite{Beisert:2003tq,Beisert:2003yb,Beisert:2003jj,Beisert:2004ry}.

\medskip
These investigations suggested that ${\cal N}=4$ SYM could admit an all loop solution for the renormalization flow 
of its composite operators. The reason for this peculiar {\em internal} integrability is the strong constraint 
imposed by the unbroken superconformal symmetry. Also, in agreement with AdS/CFT duality, the symmetry is further enhanced at tree level 
to the higher spin algebra $\mathfrak{hs}(2,2|4)$. Its multiplets obey intricate recombination rules under the interacting 
subalgebra $\mathfrak{psu}(2,2|4)$~\cite{Beisert:2004di}.

\medskip
The analysis of the dilatation operators was further extended in~\cite{Beisert:2003ys}, computing 
the three loop dilatation operator in the $\mathfrak{su}(2|3)$ subsector of the theory.
This remarkable achievement has been obtained algebraically by exploiting the superconformal invariance instead of direct Feynman diagram calculations. 
The  $\mathfrak{su}(2|3)$ sector includes closed smaller subsectors like the bosonic $\mathfrak{su}(2)$ and the fermionic $\mathfrak{su}(1|1)$,
but not the non-compact $\mathfrak{sl}(2)$ containing the twist operators.

\medskip
As is common in the context of integrable models, the basic object is not the Hamiltonian, but the $S$-matrix for elementary excitations.
Integrability means that scattering processes are elastic and factorized. To a large extent, the two body scattering matrix allows to reconstruct 
the full dynamics. Once we know the $S$-matrix, we can write down Bethe Ansatz equations that determine the 
allowed quantum numbers and momenta of the elementary excitations. From the solutions of these equations we extract the spectrum
of the integrable Hamiltonian, {\em i.e.} the anomalous dimensions in the gauge theory.

\medskip
Therefore, as soon as integrability was recognized as an essential feature of ${\cal N}=4$ SYM, a new investigation 
route started with the aim of computing higher order Bethe Ansatz equations bypassing the explicit construction of the dilatation 
operator. For instance, a remarkable result was obtained in~\cite{Serban:2004jf} where the three loop Bethe Ansatz equations were determined 
in the $\mathfrak{su}(2)\subset\mathfrak{su}(2|3)$ bosonic sector. 

\medskip
Here and in the following, the Bethe Ansatz equations that we shall discuss will always be {\em asymptotic}. 
This means that their perturbative expansion predicts the correct anomalous dimension
of the operator ${\cal O}$ up to a {\em finite} loop order ${\cal O}(g^{2L})$. The number $L$ depends in a controlled way on the number of 
elementary fields in ${\cal O}$. The dilatation operator, seen  as a spin chain Hamiltonian,
contains long-range interactions up to a distance growing with the perturbative order. 
The number of fields in ${\cal O}$ determines the length of the spin chain
and, beyond a certain order, wrapping problems appear invalidating the Bethe Ansatz equations.

\medskip
Due the simplicity of the $\mathfrak{su}(2)$ sector, it was possible to compute the 
three loop dilatation operator and its Bethe Ansatz equations at the five loops level~\cite{Beisert:2004hm}.
This remarkable result was also confirmed for many operators by standard Feynman diagram calculations 
at three loops~\cite{Eden:2004ua}.
The Bethe Ansatz and $S$-matrix approach culminated in the work~\cite{Staudacher:2004tk}.
A clever combination of gauge/string theory arguments allowed to conjecture the three loop Bethe Ansatz 
equations for the other rank-1 closed subsectors $\mathfrak{su}(1|1)$ and $\mathfrak{sl}(2)$ without 
computing the dilatation operator.

\medskip
From our point of view, the paper~\cite{Staudacher:2004tk} is crucial because it started the investigation of the
non-compact $\mathfrak{sl}(2)$ sector. As a check of the proposed Bethe Ansatz equations, a few explicit two loops
calculation have been performed in~\cite{Eden:2005bt} using  superspace perturbation theory. 
Also, the two loop dilatation operator has been constructed algebraically in the $\mathfrak{su}(1,1|2)\supset \mathfrak{sl}(2)$
sector~\cite{Zwiebel:2005er}. In all cases, the Bethe equations of~\cite{Staudacher:2004tk} were fully confirmed.

\medskip
The three loop $S$-matrix in the $\mathfrak{sl}(2)$ sector is an important bridge toward  QCD.
Based on three loop QCD calculations~\cite{Moch:2004pa} and inspired by one and two loop results in ${\cal N}=4$ SYM~\cite{Dolan:2000ut,Arutyunov:2001mh},
Kotikov, Lipatov, Onishchenko and Velizhanin (KLOV) conjectured a three loop prediction for the anomalous dimension of  ${\cal N}=4$ 
twist-2 superconformal operators at generic spin in~\cite{Kotikov:2004er}. The prediction is based on what is now called
the {\em maximum transcendentality principle}. 

\medskip
This prediction is perfectly matched by the perturbative expansion 
of the $\mathfrak{sl}(2)$ Bethe Ansatz equations. Notice that at twist-2 all conformal operators fall in a single supermultiplet
(see~\cite{Belitsky:2003sh,Belitsky:2005gr,Beisert:2002tn}). With more complicated operators this is not necessarily true.
For instance, the paper~\cite{Belitsky:2005bu} studies the two loop dilatation in ${\cal N}=1, 2, 4$ SYM
for Wilson operators with 3 quark/gaugino fields and derivatives. This is not the $\mathfrak{sl}(2)$ sector
that we are discussing and that is built with the holomorphic scalar fields of ${\cal N}=4$ SYM.

\medskip
As a technical remark, the agreement between~\cite{Staudacher:2004tk} and the KLOV prediction is somewhat beyond the regime of applicability
of the Bethe equations as it would follow from wrapping considerations. This subtle point is related to superconformal
symmetry as explained in the later works~\cite{Beisert:2005fw,Beisert:2005tm}. Wrapping problems are expected to appear at 
$L+2$ loop order for twist-$L$ operators. 

\medskip
Soon, it was realized that the wrapping barrier could be overcome by considering the large spin 
limit of anomalous dimensions. In this limit, the leading contribution to the smallest anomalous dimension 
scales logarithmically with the spin $s$ and is universal with respect to twist~\cite{Belitsky:2006en}.
The coefficient of $\log s$ is a non-trivial function of the coupling, the {\em cusp} anomalous
dimension (a.k.a. scaling function)~\cite{Korchemsky:1988si,Korchemsky:1992xv,Polyakov:1980ca}.

\medskip
Since wrapping problems do not affect the calculation of the scaling function, a possibility opened for its 
all-loop calculation from the Bethe Ansatz equations. The first attempt in this direction is described in~\cite{Eden:2006rx}. 
In that work, the  $\mathfrak{sl}(2)$ Bethe equations are put on a more solid basis,
by a field theoretical calculation of the two loop dilatation operator. This calculation confirmed that it is diagonalized by the 
Bethe Ansatz equations of~\cite{Staudacher:2004tk}. Then, an integral equation was proposed to determine at all loops
the scaling function, the Eden-Staudacher (ES) equation.
As a matter of fact, the four loop ES prediction violates the maximum transcendentality principle. 
This anomalous fact was soon related to a missing crucial piece in the Bethe Ansatz analysis, namely the 
{\em dressing factor}~\cite{Beisert:2005wv}.
It is an Abelian phase which is not constrained by superconformal symmetry and enters the perturbative expansion of 
anomalous dimensions precisely at four loops. 

\medskip
Although, at first, the appearance of the dressing factor is rather discouraging,  AdS/CFT duality comes to rescue.
The ${\cal N}=4$ SYM theory has a $\ads$ string counterpart which is also classically integrable (and to some extent also at the quantum level) 
and where 
an all-order strong coupling expansion for the dressing is known. This is made possible by a combination of 
semiclassical string calculations and string integrability considerations, like crossing 
symmetry~\cite{Beisert:2006ib,Arutyunov:2004vx,Beisert:2005cw,Hernandez:2006tk,Freyhult:2006vr,Janik:2006dc}. 

\medskip
By an impressive insight, Beisert, Eden and Staudacher proposed in~\cite{Beisert:2006ez} an all-order weak-coupling expansion of the dressing
phase. The modified integral equation for the scaling function now predicted a four loop contribution in analytical form 
in agreement with the KLOV maximum transcendentality principle.
In a synchronized but independent fashion, a remarkable four-loop Feynman diagram calculation appeared~\cite{Bern:2006ew}
providing a numerical expression for this contribution in full agreement with the result of~\cite{Beisert:2006ez}.

\medskip
More recently, there have been additional important developments aimed at a deeper understanding of the 
dressing factor~\cite{Rej:2007vm,Sakai:2007rk,Gromov:2007cd,Beisert:2007ds}. In this paper, we shall dwell on
the results of~\cite{Beisert:2006ez} to compute certain 4 loop contributions.

\medskip
A complementary approach to integrable systems (closely related to the Bethe Ansatz) is based on the Baxter 
$Q$-operator~\cite{Bax72}. In this framework, the two loop dilatation operator in ${\cal N}=2, 4$ SYM
for Wilson operators with scalars and derivatives (our $\mathfrak{sl}(2)$ sector) as well as the two loop Baxter 
operator are computed in~\cite{Belitsky:2006av}. The all-loop asymptotic generalization of the Baxter equation
in the same sector also  appeared in~\cite{Belitsky:2006wg}.
All-loop extensions to the larger $\mathfrak{sl}(2|1)$ sector have been recently published in~\cite{Belitsky:2006cp,Belitsky:2006cp}.

\medskip
In this paper, we keep working in the $\mathfrak{sl}(2)$ sector of ${\cal N}=4$ SYM and study twist-3 operators extracting from the Bethe 
Ansatz equations the (lowest) anomalous dimension.
For these operators, the Bethe Ansatz equations plus dressing are expected to be  reliable up to 4 loops, even at finite spin.
We shall show that, for even spin, the minimal anomalous dimension is associated with an unpaired operator and the 
non-dressing part of it has rational contributions up to 4 loops. 

We shall be able to provide closed expressions for the various perturbative
contributions at {\em finite spin $s$}, including the dressing part at four loops. As a non trivial check, 
we shall recover the universal four loop cusp anomalous dimension.

\medskip
The detailed plan of the paper is as follows. Sec.~(\ref{sec:Intro}) is devoted to a brief 
overview of the main integrability facts in ${\cal N}=4$ SYM with some discussion of the links with QCD
integrability. Sec.~(\ref{sec:sl2}) describes the relevant Bethe equations.
Sec.~(\ref{sec:twist3}) is devoted to the detailed presentation of our analysis on the twist-3 operators.
A self-contained Appendix contains some technical information about nested harmonic sums, which are the basic 
element entering our proposed closed formulae for anomalous dimensions at finite spin.

\section{The $\mathfrak{sl}(2)$ sector of ${\cal N}=4$ SYM}
\label{sec:sl2}

\subsection{The Bethe Ansatz equations}

The $\mathfrak{sl}(2)$ sector of planar ${\cal N}=4$ SYM contains single trace states which are linear combinations
of the basic operators
\be 
\label{eq:sl2states}
\mbox{Tr}\left\{ \left({\cal D}^{s_1}\,Z\right)\ \cdots
\left({\cal D}^{s_L}\,Z\right) \right\},\quad s_1 + \cdots + s_L = s,
\ee 
where $Z$ is one of the three complex scalar fields and ${\cal D}$ is a light-cone covariant
derivative. The numbers $\{s_i\}$ are non-negative integers and $s$ is the total spin. The number $L$ of $Z$
fields is the twist of the operator, {\em i.e.} the classical dimension minus the spin.
The subsector of states with fixed spin and twist is perturbatively closed under renormalization mixing.

\medskip
At one-loop, the dilatation operator in this sector maps to an integrable spin chain with $L$ spins transforming
according to the $s=-1$ infinite dimensional $\mathfrak{sl}(2)$ representation~\cite{Beisert:2003yb}.
Beyond one loop, the work~\cite{Staudacher:2004tk} proposed asymptotic all-order Bethe Ansatz equations.
As a check, a few explicit two loops calculation have been performed in~\cite{Eden:2005bt} using  superspace perturbation theory. 
Also, the two loop dilatation operator has been constructed algebraically in the $\mathfrak{su}(1,1|2)\supset \mathfrak{sl}(2)$
sector~\cite{Zwiebel:2005er}. It has also been evaluated in the $\mathfrak{sl}(2)$ by Feynman diagram calculations in~\cite{Eden:2006rx}.
In all cases, the Bethe equations of~\cite{Staudacher:2004tk} were confirmed.
Wrapping problems are delayed by supersymmetry and appear at $L+2$ loop order for twist-$L$ operators~\cite{Beisert:2005fw,Beisert:2005tm}
(also, N. Beisert and M. Staudacher, private communication).

\medskip
The anomalous dimensions of scaling combinations of states \refeq{eq:sl2states} are the eigenvalues 
$\gamma_L(s; g)$ of the dilatation operator/integrable Hamiltonian where our definition for the planar coupling is 
\be
g^2 = \frac{g_{\rm YM}^2\,N}{8\,\pi^2},
\ee 
where $N$ is the number of colors. The coupling $g$ is kept fixed as $N\to\infty$.
These anomalous dimensions $\gamma_L(s; g)$  are obtained by solving perturbatively 
the Bethe Ansatz equations, provided we are in the wrapping-free cases.

\medskip
The Bethe Ansatz equations determine $s$ real Bethe roots $\{u_k\}_{1\le k \le s}$ and read
\be 
\label{eq:basl2} 
\left(\frac{x^+_k}{x^-_k}\right)^L =
\prod_{j\neq k}^s\frac{x^-_k-x^+_j}{x^+_k-x^-_j} \frac{\displaystyle 1-\frac{g^2}{2\,x^+_k
x^-_j}}{\displaystyle 1-\frac{g^2}{2\,x^-_k x^+_j}}, \qquad x^\pm_k =
x\left(u_k\pm\frac{i}{2}\right), 
\ee 
where we have defined the maps 
\be x(u) = \frac{u}{2}
\left(1+\sqrt{1-\frac{2\,g^2}{u^2}}\right),\qquad u(x) = x + \frac{g^2}{2\,x}. 
\ee 
The relevant solutions of \refeq{eq:basl2} are those obeying the constraint
$\prod_{k=1}^s (x^+_k/x^-_k) = 1$ needed to project onto cyclic states associated with single-trace operators in the gauge theory.
Given a solution of Bethe Ansatz equations, we obtain the anomalous dimension from the formula
\be 
\gamma_L(s; g) = g^2\sum_{i=1}^s\left(\frac{i}{x^+_i}-\frac{i}{x^-_i}\right). 
\ee 

\bigskip
Often, the Bethe Ansatz equations are presented in a different, equivalent form, more suitable for our later discussion.
We introduce the Bethe momenta $\{p_i\}_{1\le i\le s}$ which are related to the Bethe roots by the transformations
\ba
p(u) &=& -i \log\frac{x^+(u)}{x^-(u)}, \\
u(p) &=& \frac{1}{2}\cot\frac{p}{2}\sqrt{1+8\,g^2\sin^2\frac{p}{2}}.
\ea
The $x^\pm$ combinations appearing in the Bethe equations have the following expression in terms of the $p_i$,
\be
x^\pm(p) = \frac{\displaystyle e^{\pm\frac{ip}{2}}}{4\sin\frac{p}{2}}\left(1+\sqrt{1+8\,g^2\sin^2\frac{p}{2}}\right).
\ee
Finally, the anomalous dimension reads
\be
\gamma_L(s; g) = \sum_{k=1}^s \left(\sqrt{1+8\,g^2\sin^2\frac{p_k}{2}}-1\right).
\ee
Notice that in this form, the anomalous dimension is obtained as a sum of single quasi-particle energies with a definite 
dispersion relation.

\medskip
The Bethe equations admit several solutions that describe different scaling operators. The different solutions
are selected by taking the logarithm of the Bethe equations and choosing a determination for the branch. This is known as
choosing the mode/Bethe numbers of the state.
In the following analysis we shall study the minimal anomalous dimension, {\em i.e.} the ground state of the 
integrable spin chain, for reasons that shall become
clear in the discussion. The solution describing the ground state has known mode numbers at any twist~\cite{Eden:2006rx}.

\medskip
Starting from the 4-loop level, the Bethe equations Eqs.~(\ref{eq:basl2}) must be modified including a universal
Abelian dressing phase. In other words, beyond three loops, the correct equations are 
\be \label{eq:basldress}
\left(\frac{x^+_k}{x^-_k}\right)^L = \prod_{j\neq k}^s\frac{x^-_k-x^+_j}{x^+_k-x^-_j}
\frac{\displaystyle 1-\frac{g^2}{ x^+_k x^-_j}}{\displaystyle 1-\frac{g^2}{ x^-_k x^+_j}}\
e^{2\,i\,\vartheta_{kj}},
\ee
with $\vartheta_{kj} = {\cal O}(g^6)$. The general perturbative expansion of the dressing phase
takes the form 
\be 
\label{eq:dressing}
\vartheta_{kj} = \sum_{r\ge 2}\sum_{\nu\ge 0}\sum_{\mu\ge \nu}
\left(\frac{g^2}{2}\right)^{r+\nu+\mu} \beta_{r, r+1+2\nu}^{(r+\nu+\mu)}\left[q_r(p_k)
q_{r+1+2\nu}(p_j) - (k\leftrightarrow j)\right], 
\ee 
where the higher order charges $q_r(p)$ have the expression~\cite{Beisert:2004hm}
\be q_r(p)
=\frac{2\sin\left(\frac{r-1}{2}p\right)}{r-1}\left(\frac{\sqrt{1+8\,g^2\sin^2\frac{p}{2}}-1}{2\,g^2\,\sin\frac{p}{2}}\right)^{r-1},
\ee 
with the non trivial coefficient being $\beta^{(3)}_{2,3}\neq 0$.

\medskip
The proposed formula for the coefficients $\beta_{a,b}^{(c)}$ is given in~\cite{Beisert:2006ez} (see also \cite{Gomez:2006mf,Kotikov:2006ts}
for related explorations) and reads
\be 
\beta_{r, r+1+2\nu}^{(r+\nu+\mu)} =
2(-1)^{r+\mu+1}\frac{(r-1)(r+2\nu)}{2\mu+1}\binom{2\mu+1}{\mu-r-\nu+1}\binom{2\mu+1}{\mu-\nu}\zeta(2\mu+1),
\ee 
and zero if $\mu-r-\nu+1 < 0$.
In particular, for the leading order 4-loop correction we have
\ba
2\,\vartheta_{kj} &=& \beta\,(q_2^{(0)}(p_k)\,q_3^{(0)}(p_j)-q_2^{(0)}(p_j)\,q_3^{(0)}(p_k))\,g^6 + {\cal O}(g^8), \\
q_2^{(0)}(p) &=& 4\,\sin^2\frac{p}{2}, \\
q_3^{(0)}(p) &=& 4\,\sin^2\frac{p}{2}\,\sin p,
\ea
and
\be
\beta^{(3)}_{2,3} = 4\,\beta,\qquad \beta = \zeta_3.
\ee

\subsection{Extracting the perturbative anomalous dimensions, the twist-2 case}

Let us fix the twist. For a given spin, we can start by solving numerically the one-loop Bethe Ansatz equations
(see later for more analytical information on this step).
Then, it is straightforward to work out the perturbative expansion of the all-order equations. In this way we obtain numerical values for the various loop contributions to 
\be
\gamma_L(s; g) = \sum_{n\ge 1}\gamma_L^{(n)}(s)\,g^{2\,n} .
\ee 
If these coefficients are rational, we can identify them unambiguously by working with a very large number of digits. 

\medskip
Actually, in many integrable systems, there is an alternative approach based on the $Q$ Baxter operator~\cite{Bax72}. One build a polynomial $Q(u)$
whose roots are the Bethe roots. This polynomial satisfies a recurrence equation that determines it completely, once the quantum numbers
of the desired state are chosen. From the (analytical) knowledge of $Q(u)$, it is possible to obtain the Hamiltonian eigenvalues without resorting to
finite precision numerical calculations.

\medskip
At twist-2, the one loop Baxter operator is known in closed form and the Baxter equation is also known up to three loops. The Bethe Ansatz
and Baxter method of course agree as we checked up to spin $s=68$. The resulting anomalous dimensions agrees with the
KLOV three loop prediction
\ba
\gamma_{2, s}^{(1)} &=& 4\, S_1\, ,  \\
\gamma_{2, s}^{(2)} &=&-4\,\Big( S_{3} + S_{-3}  -
2\,S_{-2,1} + 2\,S_1\,\big(S_{2} + S_{-2}\big) \Big)\, ,  
\nonumber \\
\gamma_{2, s}^{(3)} &=& -8 \Big( 2\,S_{-3}\,S_2 -S_5 -
2\,S_{-2}\,S_3 - 3\,S_{-5}  +24\,S_{-2,1,1,1}\nonumber\\
&&~~~~~~+ 
6\,\big(S_{-4,1} + S_{-3,2} + S_{-2,3}\big)
- 12\,\big(S_{-3,1,1} + S_{-2,1,2} + S_{-2,2,1}\big)\nonumber \\
&&~~~~~~-
\big(S_2 + 2\,S_1^2\big) 
\big( 3 \,S_{-3} + S_3 - 2\, S_{-2,1}\big)
- S_1\,\big(8\,S_{-4} + S_{-2}^2\nonumber \\
&&~~~~~~+ 
4\,S_2\,S_{-2} +
2\,S_2^2 + 3\,S_4 - 12\, S_{-3,1} - 10\, S_{-2,2} 
+ 16\, S_{-2,1,1}\big)
\Big)\, , \nonumber
\ea
In the above expressions, all the $S$ functions are harmonic functions (see Appendix~A) evaluated at the argument $s$, the spin
\be
S_\mathbf{a} \equiv S_\mathbf{a}(s).
\ee
This check is not surprising and extend the checks already performed in~\cite{Staudacher:2004tk}.

\section{The $\mathfrak{sl}(2)$ sector at twist-3}
\label{sec:twist3}

The same approach can be applied to the twist-3 case. Here, the one loop anomalous dimension $\gamma_3^{(1)}(s)$ is 
known from the analysis of~\cite{Beisert:2003jj} (see also some conjectures in~\cite{Beisert:2004di}).
In particular, for even spin, the ground state of the integrable Hamiltonian is an unpaired state. 

\medskip
At twist-3, the Bethe equations are believed to be reliable up to four loops. This is beyond the range of applicability of 
the known Baxter equation that stops at three loops. 
So, we have computed $\gamma_3^{(1,2,3)}(s)$ by both methods up to $s=68$ and $\gamma_3^{(4)}(s)$
in the same range by the Bethe equations only.

\medskip
This means that we have computed $\gamma_3^{(n)}(s)$ for $n=1,2,3,4$ and $s<70$ in {\em analytical form} as rational 
numbers plus, at four loop, a transcendental dressing contribution. The limit $s<70$ is fixed by computer limitations
that appear when the result is sought in analytical form. At $s=68$ and four loops we need several thousands of digits.
Of course, if a numerical result is enough, it is possible to push quite further the calculation as we shall discuss later.

\medskip
In the following sections, we shall first present new exact results for the twist-3 Baxter operator at one loop, as well as 
all the details of the Baxter three loop calculation. 

\subsection{The one-loop Baxter operator at twist-3}

We follow the notation of~\cite{Belitsky:2006en}.
The one-loop Baxter equation at twist-3 reads
\be
\left(u+\frac{i}{2}\right)^3\,Q_3\left(u+i\right) + 
\left(u-\frac{i}{2}\right)^3\,Q_3\left(u-i\right) = t_3(u)\,Q_3(u),
\ee
where
\ba
t_3(u) &=& 2\,u^3+q_2\,u+q_3,\\
q_2 &=& -\left(s+\frac{3}{2}\right)\,\left(s + \frac{1}{2}\right)-\frac{3}{4}.
\ea
Here, $s$ is the (even) spin. The Baxter operator $Q(u)$ is a polynomial of degree $s$
\be
Q_3(u) = \sum_{n=0}^s a_n\,u^n,
\ee
whose roots are the Bethe roots. Replacing $Q_3$ in the Baxter equation we obtain a homogeneous linear problem in the coefficients $\{a_n\}$. The 
parameter $q_3$ is a quantum number. It appears in the linear problem as an eigenvalue. For the lowest state, $q_3=0$. 
Also, the Baxter polynomial turns out to be even under $u\to -u$.

\medskip
The eigenvector associated with $q_3=0$ determines $\{a_n\}$ and hence $Q_3$ up to an irrelevant  scaling factor.
Given the roots $\{u_n\}$ of the Baxter polynomial 
\be
Q_3(u) = {\cal N}\cdot\prod_{n=1}^s (u-u_n),
\ee
we can compute the one-loop anomalous dimension by the formula
\be
\gamma_3^{(1)}(s) = \sum_{n=1}^s \frac{1}{u_n^2+1/4} = 2\,Q'_3\left(-\frac{i}{2}\right).
\ee
Notice that the last expression does not require the knowledge of $\{u_n\}$, but just $Q_3(u)$.

\medskip
For example, at $s=2$ we find 
\be
Q_3(u)\sim (2\,u+1)(2\,u-1),\qquad u = \pm \frac{1}{2},\qquad \gamma_3^{(1)}(2) = 4.
\ee
For $s=4$ we find 
\be
Q_3(u)\sim 11-72\,u^2+48\,u^4,\qquad u = \pm \sqrt{\frac{3}{4}\pm\frac{1}{\sqrt{3}}},\qquad \gamma_3^{(1)}(4) = 6.
\ee

\noindent
Now, with some trial and error and guided by a comment about Wilson polynomials in~\cite{Belitsky:2006en},
we found the following closed solution to the Baxter equation
\be
\label{eq:bax1}
Q_3(u) = {\cal N}\,{}_4\,F_3\left(\begin{array}{c}
\displaystyle -\frac{s}{2}, \frac{s}{2}+1, \frac{1}{2}+i\,u, \frac{1}{2}-i\,u \\
1, 1, 1
\end{array}
; 1\right).
\ee
Then, one obtains
\be
\gamma_3^{(1)}(s) = 2\,Q'_3\left(-\frac{i}{2}\right) = s\,\left(\frac{s}{2}+1\right)
\,{}_4\,F_3\left(\begin{array}{c}
\displaystyle 1, 1, 1-\frac{s}{2}, 2+\frac{s}{2} \\
2, 2, 2
\end{array}
; 1\right) = \cdots = 4\,S_1\left(\frac{s}{2}\right).
\ee
This result is new and is in perfect agreement with the empirical conjecture of~\cite{Beisert:2003jj}.

\subsection{The three loop Baxter equation at twist-3}

The Baxter equation at twist-3 valid up to three loops is described in~\cite{Belitsky:2006av}. Let $Q(u)$ be the Baxter polynomial. 
For simplicity, we omit the twist label. $Q$ can be loop expanded in the form 
\be
Q(u) = Q^{(0)}(u) + g^2\,Q^{(1)}(u) + g^4\, Q^{(2)}(u) + \cdots .
\ee
The polynomial $Q^{(0)}(u)$ has degree $s$ and is $Q^{(0)} = Q_3$ defined in Eq.~(\ref{eq:bax1}). The polynomials $Q^{(k)}(u)$, $k=1, 2$ have
degree $s-2$ and, for the ground state, are also even under $u\to -u$. Hence, they have the general form 
\be
Q^{(k)}(u) = \sum_{n=0}^{s/2-1} c_n^{(k)}\,u^{2n}.
\ee
The three loop Baxter equation reads
\be
\Delta_+\left[x\left(u+\frac{i}{2}\right)\right]\,Q(u+i) + \Delta_-\left[x\left(u-\frac{i}{2}\right)\right]\,Q(u-i) = t(u)\,Q(u),
\ee
where we have defined again
\be
x(u) = \frac{u}{2}\left(1+\sqrt{1-\frac{2\,g^2}{u^2}}\right),
\ee
and 
\ba
\Delta_\pm(x) &=& x^3\,\exp\left\{-\frac{g^2}{x}\,(\log Q(z))'_{z=\pm\frac{i}{2}} \right. \\
&& \left. 
-\frac{g^4}{4\,x^2}\left[
(\log Q(z))''_{z=\pm\frac{i}{2}} + x\,(\log Q(z))'''_{z=\pm\frac{i}{2}}
\right]
\right\}. \nonumber \\
t(u) &=& \sqrt{\Delta_+(x(u))\,\Delta_-(x(u))}\left[2 + \sum_{n=2}^\infty \frac{q_n(g)}{(x(u))^n}\right].
\ea
The charges $q_n$ have a weak-coupling expansion that for the ground state is
\ba
q_2(g) &=& -\left(s+\frac{3}{2}\right)\,\left(s + \frac{1}{2}\right)-\frac{3}{4} + \sum_{n=1}^\infty q_{2, n}\,g^{2\,n}, \\
q_3(g) &\equiv& 0, \\
q_4(g) &=& \sum_{n=1}^\infty q_{4, n}\,g^{2\,n}, \\
q_n(g) &=& 0,\ \ n>4 .
\ea
Given the (unique polynomial) solution $Q(u)$ of the Baxter equation, the 3-loop anomalous dimension is given by the formula
\be
\gamma_3(s; g) = 2\,i\,\left\{
g^2\,(\log Q(u))'_{u=\frac{i}{2}} + \frac{g^4}{4}\,(\log Q(u))'''_{u=\frac{i}{2}}
 + \frac{g^6}{48}\,(\log Q(u))'''''_{u=\frac{i}{2}} + {\cal O}(g^8)\right\},
\ee
where we exploited the parity invariance of $Q$. Notice that the above formula must be further expanded in $g^2$ since the polynomial $Q$
contains the coupling $g$.

\medskip
We did not find an analytical formula for the higher Baxter polynomials $Q^{(k)}$, $k=2,3$. However, it is straightforward to compute them for 
each value of the spin $s$. 

\noindent
For example, at $s=2$ we find
\ba
Q &=& \frac{1}{4}(4\,u^2-1)-\frac{5}{4}\,g^2 + 0\cdot g^4, \\
\gamma_3(2) &=& 4\,g^2-6\,g^4+17\,g^6 +\cdots .
\ea
At $s=4$, 
\ba
Q &=& \frac{1}{48}(48\,u^4-72\,u^2+11)-\frac{g^2}{24}(108\,u^2-47) + \\
&& + \frac{g^4}{16}(36\,u^2+53), \nonumber \\
\gamma_3(4) &=& 6\,g^2-\frac{39}{4}\,g^4+\frac{957}{32}\,g^6 + \cdots .
\ea
At $s=6$,
\ba
Q &=& \frac{1}{320}(320\,u^6-1520\,u^4+1292\,u^2-153) + \\
&& -\frac{g^2}{800}(8400\,u^4-17928\,u^2+4357) + \nonumber\\
&& +  \frac{3\,g^4}{64000}(178000\,u^4+325448\,u^2-359587), \nonumber \\
\gamma_3(6) &=& \frac{22}{3}\,g^2-\frac{443}{36}\,g^4+\frac{303115}{7776}\,g^6 +\cdots .
\ea

\subsection{The proposed four loop anomalous dimension at twist-3}

Using the methods described in the previous sections, we obtained the coefficients
$\gamma_3^{(n)}(s)$ as rational numbers for even $s<70$, plus a transcendental term at four loop
coming from the dressing to be discussed below, separately.

\medskip
Is it possible to predict a closed formula
for the coefficients as functions of $s$ ? Certainly, one needs some inspiration to convert the problem into a well-posed one.
We propose the following claim guided by several numerical explorations:

\bigskip
{\bf Twist-3 transcendentality principle:} {\em The expression of $\gamma_3^{(n)}(s)$ is obtained as a sum of transcendentality $2n-1$ (products of ) nested harmonic sums with 
positive indices  and argument $s/2$. For $n=4$ the same holds with the dressing contribution being $\zeta_3$ times a transcendentality 4 combination}.
\bigskip

Counting the number of possible harmonic sums (see Appendix~(A)), one sees that this principle combined with the $s<70$ results
fixes the three loop anomalous dimension.
One finds the remarkably simple expressions
\ba
\label{eq:three}
\gamma_{3}^{(1)} &=& 4\, S_1\, ,  \nonumber \\
\gamma_{3}^{(2)} &=& -2\,(S_3+2\,S_1\,S_2)\\
\gamma_{3}^{(3)} &=& 5\,S_5+6\,S_2\,S_3-8\,S_{3,1,1}+4\,S_{4,1}-4\,S_{2,3} + \nonumber \\
&& + S_1\,(4\,S_2^2+2\,S_4+8\,S_{3,1}), \nonumber
\ea
with all $S_a$ functions are evaluated at the argument $s/2$, i.e. half the (even) spin
\be
S_\mathbf{a} \equiv S_\mathbf{a}\left(\frac{s}{2}\right).
\ee
The argument $s/2$ is not surprising as we know from the one-loop analysis. 

\medskip
At four loops, the anomalous dimension has two contributions
\be
\gamma_{3}^{(4)} =  \gamma_{3}^{(4, \rm no \ dressing)}  + \gamma_{3}^{(4, \rm dressing)}.
\ee
The contribution from the dressing is included in the Bethe equations according to Eq.~(2.12) of~\cite{Beisert:2006ez}. The expression for
$\gamma_{3, s}^{(4, \rm dressing)}$ is easily found by assuming that $\beta$ has transcendentality 3. The current Ansatz is $\beta=\zeta_3$.
Applying our principle, one fixes 
\be
\label{eq:fourdressing}
\gamma_3^{(4, \rm dressing)} = -8\,\beta\,S_1\,S_3.
\ee
The rational part $\gamma_{3}^{(4, \rm no \ dressing)}$ cannot be fixed by the principle since the number of independent positively indexed
with transcendentality 7 is 64 and we have only 34 spin values available. However, with some trial and error, we obtained the following {\em simple}
formula
\ba
\label{eq:four}
\gamma_{3, s}^{(4, \rm no\ dressing)} &=& 
\frac{1}{2} \,S_{7}+7 \,S_{1,6}+15 \,S_{2,5}-5 \,S_{3,4}-29 
\,S_{4,3}-21 \,S_{5,2}-5 \,S_{6,1}  \\
&& -40 \,S_{1,1,5}-32 \,S_{1,2,4}+24 
\,S_{1,3,3}+32 \,S_{1,4,2}-32 \,S_{2,1,4}+20 \,S_{2,2,3} \nonumber \\
&& +40 
\,S_{2,3,2}+4 \,S_{2,4,1}+24 \,S_{3,1,3}+44 \,S_{3,2,2}+24 
\,S_{3,3,1}+36 \,S_{4,1,2}+36 \,S_{4,2,1} \nonumber \\
&& +24 \,S_{5,1,1}+80 
\,S_{1,1,1,4}-16 \,S_{1,1,3,2}+32 \,S_{1,1,4,1}-24 \,S_{1,2,2,2}+16 
\,S_{1,2,3,1} \nonumber \\
&& -24 \,S_{1,3,1,2}-24 \,S_{1,3,2,1}-24 \,S_{1,4,1,1}-24 
\,S_{2,1,2,2}+16 \,S_{2,1,3,1}-24 \,S_{2,2,1,2} \nonumber \\
&& -24 \,S_{2,2,2,1}-24 
\,S_{2,3,1,1}-24 \,S_{3,1,1,2}-24 \,S_{3,1,2,1}\nonumber \\
&& -24 \,S_{3,2,1,1}-24 
\,S_{4,1,1,1}-64 \,S_{1,1,1,3,1} . \nonumber
\ea
This formula is minimal in the sense that it is the only solution with coefficients having 2 as the largest denominator. Notice that there are
no solutions with integer coefficients.
Eqs.~(\ref{eq:three}, \ref{eq:fourdressing} and \ref{eq:four}) are the main result of this paper.

\medskip
Now, given the above formulas, one can give up the goal of obtaining $\gamma_3^{(n)}(s)$ as exact rational numbers.
Working with a moderate precision it is possible to go quite further in the spin. We computed the coefficients $\gamma_3^{(n)}(s)$
with 200 digits accuracy up to $s=300$. The result can be compared with the guessed formulae above. The equality is perfect for all digits !
This means that the proposed $\gamma_3^{(4)}$ is actually the {\em only} solution compatible with our twist-3 principle. Also, at three loops,
one could exclude possible weaker forms of the principle admitting nested harmonic sums with some negative index.

\medskip
In the next Section, we shall test these expression in the large spin limit showing that they reproduce the 
four-loop universal cusp anomalous dimension.

\subsection{The cusp anomalous dimension}

The cusp anomalous dimension $\Gamma_{\rm cusp}(g)$ is defined in the large $s$ limit of the minimal 
twist-$L$ anomalous dimension as
\ba
\gamma_L(s; g) &=& \Gamma_{\rm cusp}(g)\,\log\,s + \mbox{subleading at $s\to \infty$}, \\
\Gamma_{\rm cusp}(g) &=& \sum_{n\ge 1} \Gamma^{(n)}_{\rm cusp}\,g^{2n}.
\ea
It is expected to be twist-independent. Its all-loop perturbative expansion is generated by 
the Beisert, Eden, Staudacher equation~\cite{Beisert:2006ez}. We shall use it as a non-trivial check of our finite
spin expressions. For completeness, we mention that $\Gamma_{\rm cusp}(g)$ is quite an interesting object in the context
of AdS/CFT duality. From the dual string theory, it is known at strong coupling at leading and next-to-leading 
order~\cite{Gubser:2002tv,Frolov:2002av,Frolov:2006qe}. Checks of this strong coupling limit have been 
recently discussed in~\cite{strong}.

\medskip
The large spin expansion of nested harmonic sums can be accomplished by the methods summarized in Appendix~(A).
The calculation is straightforward for the 1, 2, 3 loop contributions. Using the results from App.~(\ref{app:zeta}), we find
\ba
\Gamma_{\rm cusp}^{(1)} &=& 4,  \\
\Gamma_{\rm cusp}^{(2)} &=& -4\,S_2(\infty),\\
\Gamma_{\rm cusp}^{(3)} &=& 4\,S_2^2(\infty)+2\,S_4(\infty)+8\,S_{3,1}(\infty). \nonumber
\ea
The exact values are
\be
S_2(\infty) = \zeta_2 = \frac{\pi^2}{6}, \quad
S_4(\infty) = \zeta_4 = \frac{\pi^4}{90},\quad
S_{3,1}(\infty) = \zeta_4 + \frac{1}{10}\,\zeta_2^2 = \frac{\pi^4}{72}.
\ee
Collecting, we find the correct result
\be
\Gamma_{\rm cusp}(g) = 4\,g^2-\frac{2\,\pi^2}{3}\,g^4 + \frac{11\,\pi^4}{45}\,g^6 + {\cal O}(g^8).
\ee
The expansion of the four loop contribution 
needs some reshuffling. An alternative, equivalent form more suitable to study the large spin expansion turns out to be 
\ba
\gamma_{3, s}^{(4, \rm no\ dressing)} &=& 
\frac{8}{3} \,(5 \,S_{4}-4 \,S_{3,1})\,\,S_1^3 + \\ 
&& + 4\, (5 \,S_{5}-10 \,S_{3,2}-14 \,S_{4,1}+16 \,S_{3,1,1})\,\,S_1^2 + \nonumber \\
&& + (32 \,S_{3}^2+8 \,S_{2} \,S_{4}+47 \,S_{6}+88 \,S_{2} \,S_{3,1}-64 \,S_{4,2}+48 \,S_{5,1}-24 \,S_{2,2,2} \nonumber \\
&& -104 \,S_{2,3,1}+120 \,S_{4,1,1}-192 \,S_{3,1,1,1})\,\,S_1 \nonumber \\
&& -2 \,S_{3} \,S_{2}^2-\,S_{5} \,S_{2}-4 \,S_{3,2} \,S_{2}+60 \,S_{4,1} \,S_{2}-120 \,S_{3,1,1} \,S_{2} \nonumber \\
&& -\frac{289}{3} \,S_{3} \,S_{4}-\frac{189 \,S_{7}}{2}-\frac{256}{3} \,S_{3} \,S_{3,1}+136 \,S_{4,3}-24
   \,S_{5,2}-32 \,S_{6,1}-64 \,S_{2,4,1} \nonumber \\
&& +64 \,S_{3,2,2}+80 \,S_{3,3,1}-80 \,S_{5,1,1}+128 \,S_{2,3,1,1}-128 \,S_{4,1,1,1}+256 \,S_{3,1,1,1,1}. \nonumber
\ea
In this form, the techniques of Appendix~(A) are enough to expand at large $s$ the four loop anomalous dimension. Including the dressing term, 
the result is 
\be
\label{eq:cusp4}
\Gamma_{\rm cusp}^{(4)} = -\frac{73\,\pi^6}{630}+4\,\zeta_3^2-8\,\beta\,\zeta_3.
\ee
With the standard choice $\beta = \zeta_3$, we obtain 
\be
\Gamma_{\rm cusp}(g) =  4\,g^2-\frac{2\,\pi^2}{3}\,g^4+\frac{11\,\pi^4}{45}\,g^6 - \left(\frac{73\,\pi^6}{630}+4\,\zeta_3^2\right)\,g^8 + \cdots.
\ee
It agrees with the analytical prediction by Beisert, Eden and Staudacher~\cite{Beisert:2006ez}
as well as with the independent numerical prediction of~\cite{Bern:2006ew} from a remarkable four loop 
Feynman diagram computation.

\medskip
Notice that Eq.~(\ref{eq:cusp4}) comes from many contributions and requires several cancellations
of higher powers of $\log\,s$. It tests rather severely our 4 loop 
proposed expression.

\medskip
The next-to-leading contributions are also interesting and shall be discussed in a forthcoming publication~\cite{forth}.
Here, for completeness, we quote some of them. The expansion has the general form
\ba
\gamma &=& \rho\,\log\,\bar s + \sum_{k=0}^\infty\frac{1}{s^k}\sum_{\ell=0}^k \rho_{k,\ell}\,\log^\ell\,\bar s, \\
\bar s &=& \frac{1}{2}\,s\,e^{\gamma_E}.
\ea
The first terms are
\ba
\rho &=& 4\,g^2-\frac{2\,\pi^2}{3}\,g^4+\frac{11\,\pi^4}{45}\,g^6 + \left(-\frac{73\,\pi^6}{630}+4\,\zeta_3^2-8\,\beta\,\zeta_3\right)\,g^8 + \cdots, \nonumber\\
\rho_{00} &=& -2\,\zeta_3\,g^4 + \left(\frac{\pi^2}{3}\,\zeta_3-\zeta_5\right)\,g^6 + \mbox{constant}\cdot g^8 + \cdots, \nonumber\\
\rho_{10} &=& 4\,g^2-\frac{2\,\pi^2}{3}\,g^4 + \left(\frac{11\,\pi^4}{45}-4\,\zeta_3\right)\,g^6 + \left(-\frac{73\,\pi^6}{630}+\frac{4\,\pi^2}{3}\,\zeta_3 + 
4\,\zeta_3^2-8\,\beta\,\zeta_3-2\,\zeta_5\right)\,g^8 + \cdots \nonumber \\
\rho_{11} &=& 0\cdot g^2+8\,g^4-\frac{8\,\pi^2}{3}\,g^6 + \frac{6\,\pi^4}{5}\,g^8 + \cdots\nonumber
\ea
Also, we computed
\ba
\rho_{22} &=& -8\,g^6+4\,\pi^2\,g^8 + \cdots\nonumber \\
\rho_{33} &=& \frac{32}{3}\,g^8 + \cdots\nonumber
\ea
The coefficient of $\log s$, {\em i.e.}  $\rho$ is precisely the cusp anomalous dimension.

\section{Conclusions}

In summary, we have considered twist-3 operators with spin $s$ in the $\mathfrak{sl}(2)$ sector of ${\cal N}=4$ SYM built out 
of three scalar fields with derivatives. We have extracted from the Bethe Ansatz equations
the exact lowest anomalous dimension $\gamma(s)$ of scaling fields for several values of even $s$.
Here, {\em exact} means in analytical form as a known coefficient at each loop, up to the four loop level where wrapping problems
invalidate the Bethe equations. 

From these results, we have been able to provide closed expressions for the spin dependence of 
$\gamma(s)$ up to the four loop level and including the contributions from the dressing phase. 
The expressions satisfy a rather simple and new {\em transcendentality principle} extending at twist-3 the KLOV idea~\cite{Kotikov:2004er}. 
As an application, we have computed the large $s$ limit of $\gamma(s)$ and checked that the four loop universal 
cusp anomalous dimension is reproduced.

The availability of the full spin dependence allows to test generic features of anomalous dimensions in 
conformally invariant planar field theories in the spirit of~\cite{Basso:2006nk,Dokshitzer:2006nm}. More detailed results
on this important topic will appear in a forthcoming publication~\cite{forth}.

From the point of view of integrability, it is very interesting to test the consequences of the complicated dressing phase Eq.~(\ref{eq:dressing})
at finite spin. Indeed, the only weak-coupling test of of the conjectured dressing phase has been the calculation of the 
cusp anomalous dimension at infinite $s$. We consider a relevant result that the leading weak-coupling contribution from dressing 
nicely fits the twist-3 transcendentality principle at finite spin. From this point of view, the choice of twist-3 has been 
crucial to push wrapping effects beyond four loops where the dressing first appears.

As a final comment, we remark that twist-3 operators in QCD built out of quarks and gluon fields 
have a well-known phenomenological relevance in polarized DIS applications~\cite{Belitsky:1999bf,Anselmino:1994gn,Abe:1998wq}.
Here, we have considered the {\em scalar} $\mathfrak{sl}(2)$ sector which is not necessarily related to the more QCD-like channels.
However, we believe that the ideas presented in this paper will also be useful in other twist-3 sectors, possibly exploiting 
supersymmetry to partially connect different channels.

\acknowledgments
We thank G. Marchesini and Yu. Dokshitzer for very useful comments. We also thank N. Beisert and  M. Staudacher for kind
discussions about wrapping and dressing effects.

\appendix

\section{Harmonic sums}

We collect in this Appendix some useful properties of (nested) Harmonic sums. General useful
references can be found in~\cite{Harmonic}.

\subsection{Definition}

For any integer $N$ and multi-index $\mathbf{a} = (a_1, \dots, a_k) \in\zz^k$, we define recursively
\ba
S_a(N) &=& \sum_{n=1}^N\frac{1}{n^a},\\
S_{a\,\mathbf{X}}(N) &=& \sum_{n=1}^N\frac{1}{n^a}\, S_{\mathbf X}(n).
\ea
In the following we shall need only the $a_i>0$ case.

\subsection{Some shuffle algebra relations}

The product of a simple $S_a$ and a nested sum $S_\mathbf{b}$ can be written
\ba
\label{eq:shuffle}
S_a\,S_{b_1,\dots, b_k} &=& S_{a, b_1, \dots, b_k} + S_{b_1, a, b_2, \dots, b_k} + \cdots + 
S_{b_1, \dots, b_k, a} \\
&& -S_{a+b_1, \dots, b_k}-S_{b_1, a+b_2, \dots, b_k} -\cdots -S_{b_1, \dots, a+b_k}.\nonumber
\ea
In particular
\be
S_a\,S_b = S_{ab} + S_{ba}-S_{a+b}.
\ee
These relations can be used to reduce sums of the form $S_{a\cdots a}$. One finds for instance
\ba
S_{aa} &=& \frac{1}{2}(S_a^2+S_{2a}), \\
S_{aaa} &=& \frac{1}{6}(S_a^3+3\,S_a\,S_{2a}+2\,S_{3a}), \\
S_{aaaa} &=& \frac{1}{24}(S_a^4+6\,S_a^2\,S_{2a}+3\,S_{2a}^2+8\,S_a\,S_{3a}+6\,S_{4a}).
\ea
\subsection{Linear relations}

Given a particular nested sum $S_\mathbf{a}$, there are linear relations between all the sums
\be
S_{\mathbf{a}'},\qquad \mathbf{a'} = \pi{\mathbf{a}},\qquad \pi = \mbox{permutation},
\ee
involving also $S$ sums with a smaller number of indices. These linear relations are obtained from the equations
(one for each permutation $\mathbf{a}'$)
\be
S_{a'_1}\,S_{a'_2,a'_3,a'_4,\dots} = \mbox{shuffle relation Eq.~(\ref{eq:shuffle})}.
\ee
For instance, taking $\mathbf{a} = (1,1,3)$ we find 
\ba
S_1\,S_{1,3} &=& S_{1,1,3}+S_{1,1,3}+S_{1,3,1}-S_{2,3}-S_{1,4}, \\
S_1\,S_{3,1} &=& S_{1,3,1}+S_{3,1,1}+S_{3,1,1}-S_{4,1}-S_{3,2}, \\
S_3\,S_{1,1} &=& S_{3,1,1}+S_{1,3,1}+S_{1,1,3}-S_{4,1}-S_{1,4}.
\ea
They can be used to obtain two of the three 3-index sums in terms of the third.

\subsection{Evaluation of $S_\mathbf{a}$ at $N=\infty$}
\label{app:zeta}

We define the generalized, finite $N$,  $\zeta$-functions
\be
Z_\mathbf{a}(N) = \sum_{N\ge n_1 > n_2 > \cdots > n_r > 0}\frac{1}{n_1^{a_1}\cdots n_r^{a_r}}.
\ee
The values of $Z$ at $N=\infty$ are the multiple zeta functions
\be
Z_\mathbf{a}(\infty) \equiv \zeta_\mathbf{a}.
\ee
Multiple zeta functions can be reduced in terms of (known) elementary zeta functions. The relations between $Z$ and $S$-sums are trivial. For instance
\ba
S_a(\infty) &=& \zeta_a, \\
S_{a, b}(\infty) &=& \zeta_{a, b} + \zeta_{a+b}, \\
S_{a, b, c}(\infty) &=& \zeta_{a, b, c} + \zeta_{a+b, c} + \zeta_{a, b+c} + \zeta_{a+b+c}.
\ea 
The general case is obtained by summing over all possible $\zeta_\mathbf{a}$ obtained by splitting
the multi-index of $S$ in order-respecting groups and taking the sum within each group. For instance
\ba
S_{a,b,c,d}(\infty) &=& 
\zeta_{a,b,c,d} + 
\zeta_{a+b,c,d} + 
\zeta_{a,b+c,d} + 
\zeta_{a,b,c+d} + 
\zeta_{a+b,c+d} + \nonumber \\
&& +
\zeta_{a+b+c,d} + 
\zeta_{a,b+c+d} + 
\zeta_{a+b+c+d}.
\ea

Some examples relevant to compute the three loop cusp anomalous dimension are the following. From 
\ba
\zeta_{3,2} &=& -\frac{11}{2}\,\zeta_5 + 3\,\zeta_2\,\zeta_3, \\
\zeta_{3,1} &=& \frac{1}{10}\,\zeta_2^2,\\
\zeta_{4,1} &=& \zeta_{3,1,1} = 2\,\zeta_5 -\zeta_2\,\zeta_3, 
\ea
we obtain 
\ba
S_{3,2}(\infty) &=& -\frac{9}{2}\,\zeta_5 + 3\,\zeta_2\,\zeta_3, \\
S_{3,1}(\infty) &=& \zeta_4 + \frac{1}{10}\,\zeta_2^2 = \frac{\pi^4}{72},\\
S_{4,1}(\infty) &=& 3\,\zeta_5 -\zeta_2\,\zeta_3, \\
S_{3, 1, 1}(\infty) &=& -\frac{1}{2}\,\zeta_5 + \zeta_2\,\zeta_3.
\ea

\subsection{Asymptotic expansions of harmonic sums with positive indices}

We first define
\be
S_a^{(p)}(N) = \sum_{n=1}^N\frac{\log^p n}{n^a}.
\ee
The following expansions hold ($B_k$ are Bernoulli's numbers)
\ba
S_1(N) &=& \log\,N+\gamma_E+\frac{1}{2\,N}-\sum_{k\ge 1}\frac{B_{2\,k}}{2\,k\,N^{2\,k}}, \\
S_a(N) &=& \zeta_a + \frac{a-2\,N-1}{2\,(a-1)\,N^a}-\frac{1}{(a-1)!}\sum_{k\ge 1}\frac{(2\,k+a-2)!\,B_{2\,k}}{(2\,k)!\,N^{2\,k+a-1}}, \quad a\in \mathbb{N}, a>1.
\ea
Taking derivatives with respect to $a$ we obtain immediately expansions for $S_a^{(p)}(N)$.

\medskip
Multiple (nested)  sums $S_\mathbf{a}$ can be evaluated as follows. Let $\mathbf{a} = (a_1, a_2, \dots, a_k)$. Suppose that the expansion of 
$S_{a_2, \dots, a_k}(N)$ is known. Its general form will always be
\be
S_{a_2, \dots, a_k} = \sum_{p,q} c_{p,q}\, \frac{\log^p N}{N^q}.
\ee
Replacing this expansion in 
\be
S_\mathbf{a} = \sum_{n=1}^N \frac{1}{n^{a_1}}\,S_{a_2,\dots, a_k}(n),
\ee
we obtain 
\be
S_\mathbf{a} = \sum_{p,q} c_{p,q} \sum_{n=1}^N \frac{\log^p n}{n^{a_1+q}}
\ee
Usually, this determines the expansion of the sum apart from the constant term. This is $S_\mathbf{a}(\infty)$ and 
can be evaluated by the methods discussed in the previous sections. Often, it is useful to reduce internal sums
of the form $S_{aaa\dots a}$ with the general formulae that we also discussed.

\subsection{Counting fixed transcendentality terms}

Let us consider a given index set $\mathbf{a} = (a_1, \dots, a_n)$ with $a_i>0$.
The various $S_{\mathbf{a}'}$ with $\mathbf{a}'$ being a permutation of $\mathbf{a}$ are not all independent due 
to the shuffle-algebra relations. The number of independent sums is 
counted by the second Witt's formula. Suppose that 
\be
\mathbf{a} = (\underbrace{a,\cdots, a}_{n_1}, \underbrace{b\cdots b}_{n_2}, \cdots ),
\ee
and let $n = n_1 + n_2 + \cdots + n_N$. 
The number of independent sums with this index set, up to permutations, is 
\be
\ell(\mathbf{a})\equiv \ell_n(n_1, \cdots, n_N) = \frac{1}{n}\sum_{d\,|\,n_i}\mu(d)\frac{(n/d)!}{(n_1/d)!\,\cdots\,(n_N/d)!},
\ee
where the M\"obius function $\mu(d)$ is 
\be
\mu(d) = \left\{\begin{array}{ll}
1, & d=1, \\
0, & \mbox{$d$ contains the factor $p^2$ with $p$ prime $>1$}, \\
(-1)^s, & \mbox{$d$ is the product of $s$ distinct primes $p_i>1$}.
\end{array}\right.
\ee
Up to transcendentality 7, the cases that occur are (all different letters stand for different numbers)
\ba
\ell(a) &=& \ell_1(1) = 1, \\
\ell(\underbrace{a\cdots a}_{n>1\ \ terms}) &=& \ell_n(n) = 0.
\ea
(This relation is well known since $S_{aaa\cdots a}$ can always be expressed in terms of product of smaller $S$)
\ba
\ell(\underbrace{a\cdots a}_{n\ \  terms}b) &=& \ell_{n+1}(1,n) = \frac{1}{n}\frac{(n+1)!}{1!\cdot n!} = 1, \\
\ell(aabb) &=& \ell_4(2,2) = \frac{1}{4}\left[\mu(1)\frac{4!}{2!\cdot 2!}+\mu(2)\frac{2!}{1!\cdot 1!}\right] = 1, \\
\ell(abc) &=& \ell_3(1,1,1) = \frac{1}{3}\frac{3!}{1!\cdot 1!\cdot 1!} = 2, \\
\ell(aaabb) &=& \ell_5(2,3) = \frac{1}{5}\frac{5!}{2!\cdot 3!} = 2, \\
\ell(aabc) &=& \ell_4(2,1,1) = \frac{1}{4}\frac{4!}{2!\cdot 1!\cdot 1!} = 3.
\ea
Now, let us begin with transcendentality 1. There is a single sum with multiplicity $\ell = 1$
\ba
\mathbf{a} &:& \ell\nonumber \\
1 &:& 1\nonumber
\ea
Hence, the number of independent positive sums with transcendentality 1 is $b_1 = 1$.

\medskip
\noindent
At transcendentality 2 we have
\ba
\mathbf{a} &:& \ell\nonumber \\
11 &:& 0\nonumber \\
2 &:& 1\nonumber
\ea
Hence, the number of independent positive sums with transcendentality 2 is $b_2 = 1$.

\medskip
\noindent
At transcendentality 3 we have
\ba
\mathbf{a} &:& \ell\nonumber \\
111 &:& 0\nonumber \\
12 &:& 1\nonumber \\
3 &:& 1\nonumber 
\ea
Hence, the number of independent positive sums with transcendentality 3 is $b_3 = 2$. From the same table 
we can also count the number $N_3$ of (products of) simple sums with total transcendentality 3. Let 
\be
R_{n,k} = \binom{n+k-1}{k},
\ee
be the number of combinations of $n$ objects in groups of $k$ with repetitions. From the table, we read
\be
N_3 = R_{b_1, 3} + b_1\,b_2 + b_3 = 4.
\ee

\medskip
\noindent
At transcendentality 4 we have
\ba
\mathbf{a} &:& \ell\nonumber \\
1111 &:& 0\nonumber \\
112 &:& 1\nonumber \\
13 &:& 1\nonumber \\
4 &:& 1\nonumber \\
22 &:& 0\nonumber 
\ea
Hence, the number of independent positive sums with transcendentality 4 is $b_4 = 3$.

\medskip
\noindent
At transcendentality 5 we have
\ba
\mathbf{a} &:& \ell\nonumber \\
11111 &:& 0\nonumber \\
1112 &:& 1\nonumber \\
113 &:& 1\nonumber \\
14 &:& 1\nonumber \\
5 &:& 1\nonumber \\
122 &:& 1\nonumber \\
23 &:& 1\nonumber 
\ea
Hence, the number of independent positive sums with transcendentality 5 is $b_5 = 6$.
From the same table 
we can also count the number $N_5$ of (products of) simple sums with total transcendentality 5. It is
\be
N_5 = R_{b_1, 5} + R_{b_1,3}\,b_2 + R_{b_1, 2}\,b_3 + b_1\,b_4 + b_5 + b_1\,R_{b_2, 2} + b_2\,b_3 = 16.
\ee

\medskip
\noindent
At transcendentality 6 we have
\be
\begin{array}{ccc}
\mathbf{a} &:& \ell \\
111111 &:& 0 \\
11112 &:& 1 \\
1113 &:& 1 \\
114 &:& 1 \\
15 &:& 1 \\
6 &:& 1 
\end{array}\qquad
\begin{array}{ccc}
\mathbf{a} &:& \ell \\
1122 &:& 1 \\
123 &:& 2 \\
24 &:& 1 \\
222 &:& 0 \\ \\ \\
\end{array} 
\ee
Hence, the number of independent positive sums with transcendentality 6 is $b_6 = 9$.
\medskip
\noindent

At transcendentality 7 we have
\be
\begin{array}{ccc}
\mathbf{a} &:& \ell \\
1111111 &:&  0\\
111112 &:&  1\\
11113 &:&  1\\
1114 &:&  1\\
115 &:&  1\\
16 &:&  1\\
7 &:&  1\\
11122 &:&  2\\
\end{array}\qquad
\begin{array}{ccc}
\mathbf{a} &:& \ell \\
1123 &:&  3\\
124 &:&  2\\
25 &:&  1\\
133 &:& 1 \\
34 &:&  1\\
223 &:&  1\\
1222 &:&  1
\end{array} 
\ee
Hence, the number of independent positive sums with transcendentality 7 is $b_7 = 18$. The same counting as before, 
taking into account the non trivial $R_{b_3, 2} = R_{2,2}=3$ (i.e. aa, ab and bb) gives
\be
N_7 = 64.
\ee
We have computed $N_k$ for the even and odd $k$ up to $k=11$ and checked that 
\be
N_k = 2^{k-1}.
\ee
The same counting can be done including sums with one or more negative indices. The number of sums $b^\pm_k$ for $k=1, \dots, 7$ is now
\be
b^\pm_k = 2, 3, 8, 18, 48, 116, 312.
\ee
Evaluating $N^\pm_k$, the general formula seems to be 
\be
N_k^\pm = 2\cdot 3^{k-1}.
\ee

\end{document}